\documentclass[aps,prl,reprint]{revtex4-2}

\usepackage{epsfig,graphics}
\usepackage{amsfonts,amssymb,amsmath} 
\usepackage{amsthm}         
\usepackage{color}
\usepackage{makecell}
\usepackage{xcolor}

\definecolor{myblue}{RGB}{0, 102, 204}

\theoremstyle{plain}

\begin{document}
\title{Orbital Angular Momentum Experiment Converting Contextuality into Nonlocality}

\author{Jianqi Sheng$^1$}

\author{Dongkai Zhang$^{1, 2}$}
\email[]{zhangdk@hqu.edu.cn}

\author{Lixiang Chen$^1$}
\email[]{chenlx@xmu.edu.cn}

\affiliation{$^1$Department of Physics, Xiamen University, Xiamen 361005, China}
\affiliation{$^2$College of Information Science and Engineering, Fujian Provincial Key Laboratory of Light Propagation and Transformation, Huaqiao University, Xiamen 361021, China}

\begin{abstract}
It was recently revealed by Cabello in a theoretical Letter [Phys. Rev. Lett. 127, 070401 (2021)] that non-locality and contextuality, as two intuitively distinctive yet both critical quantum resources, can be surprisingly connected through Bell inequalities associated with state-independent contextuality sets. It provides a general unified method capable of converting contextuality into bipartite nonlocality. However, experimental tests of the inequalities are challenging and noise-sensitive, and the requirements for the quantum states purity, dimensionality, and degree of entanglement have blocked the experimental implementation. We report a first experimental test of Cabello's inequalities from state-independent contextuality sets, by leveraging two-photon high-dimensional orbital angular momentum entangled states. Distinguishing from the standard single-particle ways, our experiment spotlights that the state-independent contextuality sets can be tested in a bipartite scenario, by which the ''compatibility'' or ''sharpness'' loopholes can be effectively avoided. Our results provide a new perspective and demonstrate the principle that contextuality, a widely used quantum resource, can be used in different physical scenarios. 
\end{abstract}

\maketitle

Quantum non-locality and contextuality are two crucial resources for quantum information processing. Non-locality describes the correlation between space-like separated measurements that any locally realistic theory cannot explain \cite{brunner2014bell}. Beyond being relevant to quantum fundamentals, this correlation has technological applications such as secure random number generation \cite{pironio2010random} and cryptography \cite{acin2007device}. Quantum contextuality describes correlations absent from classical physics similarly, that the result of a measurement depends on other jointly performed compatible measurements, but without the condition of spacelike separation \cite{budroni2022kochen}. The non-classical properties of contextual correlations have been directly connected to practical applications such as quantum communication \cite{cubitt2010improving,saha2019state,gupta2023quantum}, cryptography \cite{cabello2011hybrid,zhen2023device}, and speed-up computation \cite{raussendorf2013contextuality,delfosse2015wigner,bravyi2018quantum}. 
As the demand for quantum information processing grows, some tasks cannot be accomplished using either nonlocality or contextuality individually. This raises the question of how these two concepts are related \cite{stairs1983quantum,heywood1983nonlocality}. 
Despite the relation between non-locality and contextuality being widely investigated mathematically \cite{abramsky2011sheaf,abramsky2011cohomology,amaral2018noncontextual,abramsky2020logic} and theoretically \cite{cabello2001all,aolita2012fully,wright2023contextuality}, only very few experiments have been performed where the two notions both played important roles within a single framework \cite{liu2016nonlocality,xue2023synchronous}. 

Cabello noticed that any form of contextuality can be converted into a quantum violation of a bipartite Bell inequality in a unified way. As explained in Ref. \cite{cabello2021converting}, given a state-independent contextuality (SI-C) set, one can construct a related Bell inequality that is violated when one party chooses measurement projectors in this set and the other party chooses from the complex conjugate of the projectors. 
In previous experimental tests of contextuality, performing joint compatibility measurements on a single particle was a significant challenge \cite{budroni2022kochen}. To deal with actual experiments, a possible way is to perform nondisturbance measurements in sequence \cite{kirchmair2009state,guhne2010compatibility}. 
In experiments performed with photons, a standard measurement with the single detector is not suitable, as the photon is absorbed and no further sequence of measurements can be carried out. 
The alternative is to construct an interferometric setup for each measurement \cite{amselem2009state,lapkiewicz2011experimental}. Furthermore, one needs to identify which part of the device corresponds to each single measurement and perform additional experimental runs to quantify deviations from ideal measurements, which implies more complex procedures and introduces cumulative amounts of noise. 
In contrast, Cabello's pioneering approach makes it possible to test a SI-C set by joint measurements on spatially separated systems instead of single-particle systems. In this way, the “compatibility” \cite{szangolies2013tests,szangolies2015testing} or “sharpness” loophole \cite{vallee2024corrected,spekkens2014status,kunjwal2015kochen,krishna2017deriving} can be effectively avoided.

\begin{figure*}[!tb]
\centering
\includegraphics[width=15cm]{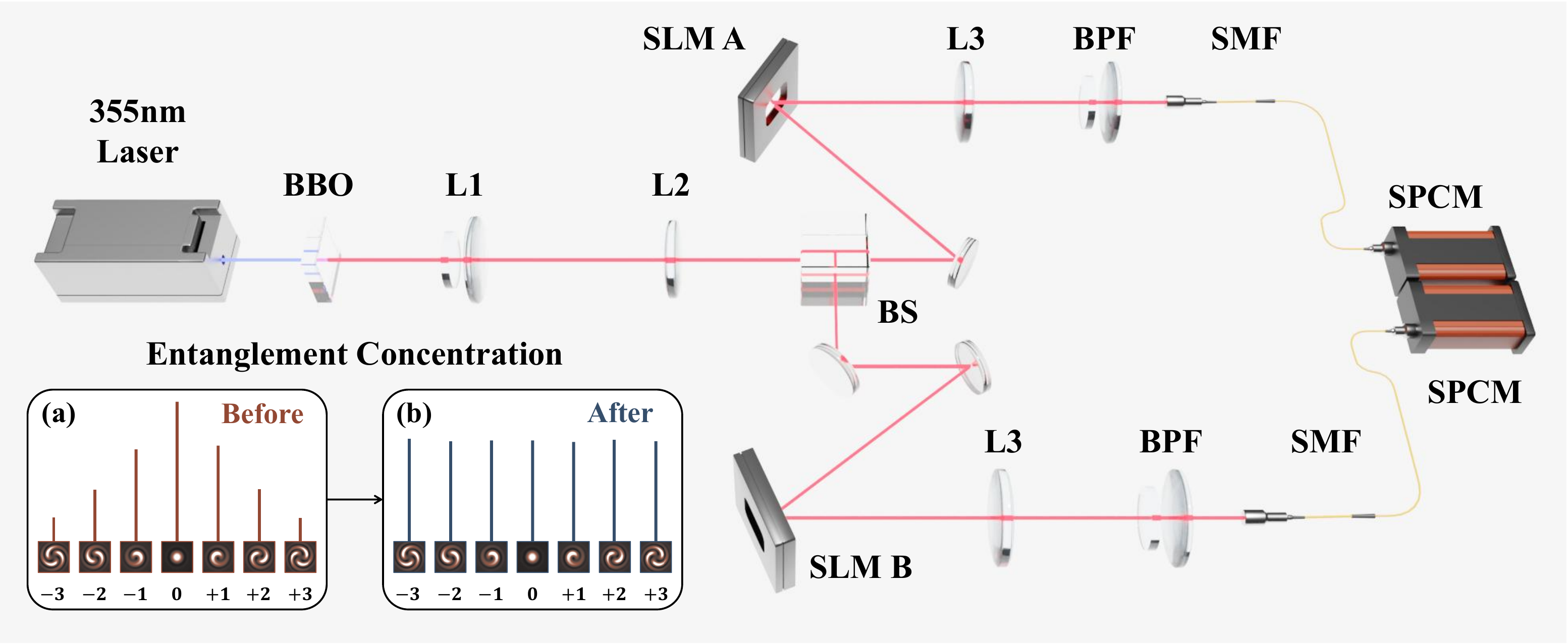}
\caption{\sf Schematic diagram of the experimental setup. BBO \(\beta\)-barium borate crystal, BS 50:50 beam splitter, SLM spatial light modulator, L lens (\(f_1=100\) mm, \(f_2=400\) mm, \(f_3=500\) mm), BPF band-pass filter, SMF single-mode fiber, SPCM single photon counting module, the outputs are connected to a coincidence circuit. The inset (a) shows the original two-photon OAM spectrum of limited spiral bandwidth before entanglement concentration, while (b) shows the maximally entangled OAM spectrum after concentration, obtained by measuring the state \(\left | \ell  \right \rangle_{A}\otimes\left | -\ell  \right \rangle_{B}\) (see Supplementary Material \cite{supp} for more details of the entanglement concentration)}.
\label{fig1}
\end{figure*}

\begin{figure*}[!tb]
\centering
\includegraphics[width=15cm]{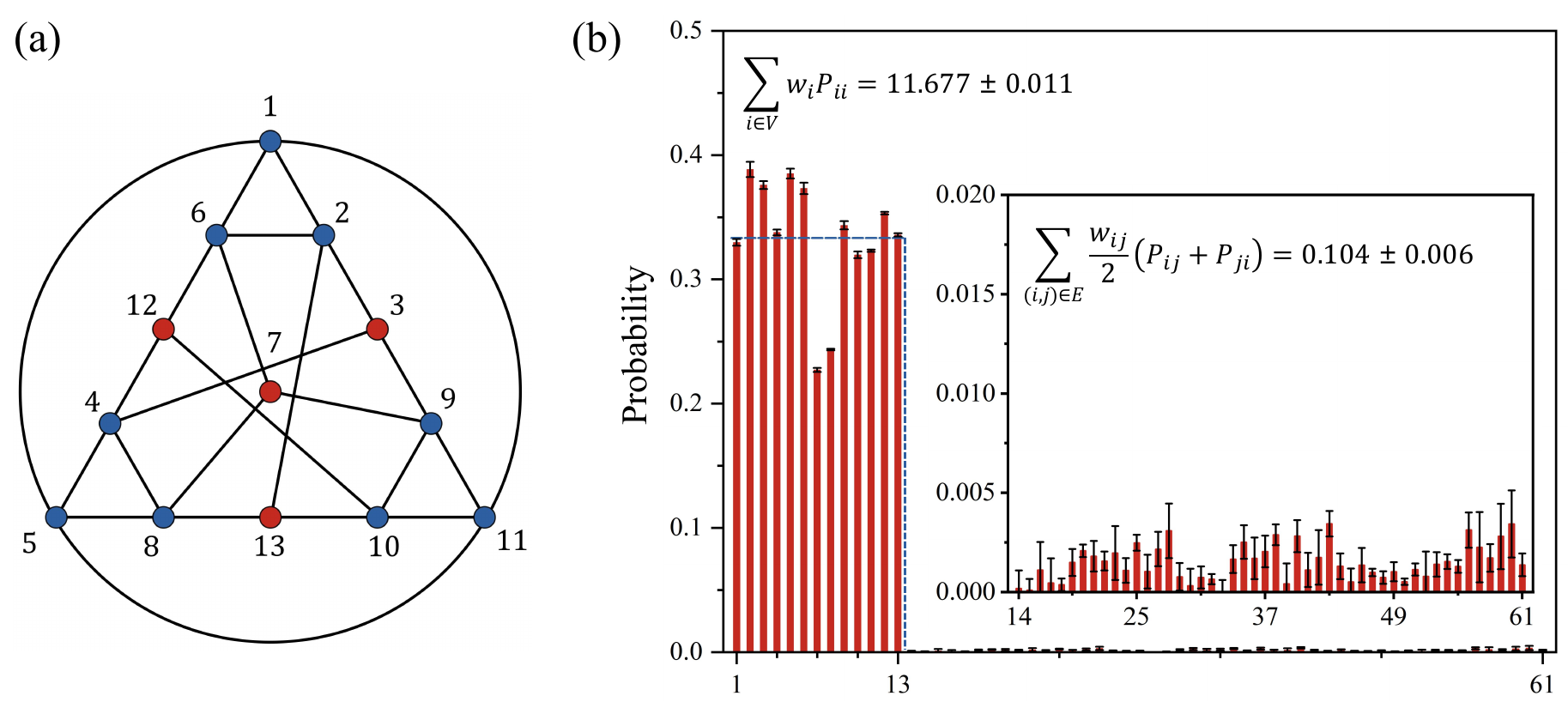}
\caption{\sf (a) Graph of orthogonality between the vectors of the YO13 set \cite{Yu2011}, the vertice with number \(i\) represents each vector \(v_i\), \( i\in \left \{ 1,\dots ,13 \right \}\). Orthogonal vectors are represented by adjacent vertices. Vertices in red have weight 2 and vertices in blue have weight 3. (b) Detailed experimental results leading to \(\beta\left(\text {YO13}\right)=11.573 \pm 0.012\) (see Supplementary Material \cite{supp} for more details). The blue dashed line is the theoretical value predicted by quantum theory for the maximally entangled state and ideal experimental apparatus.}
\label{fig2}
\end{figure*}

\begin{figure*}[!tb]
\centering
\includegraphics[width=15cm]{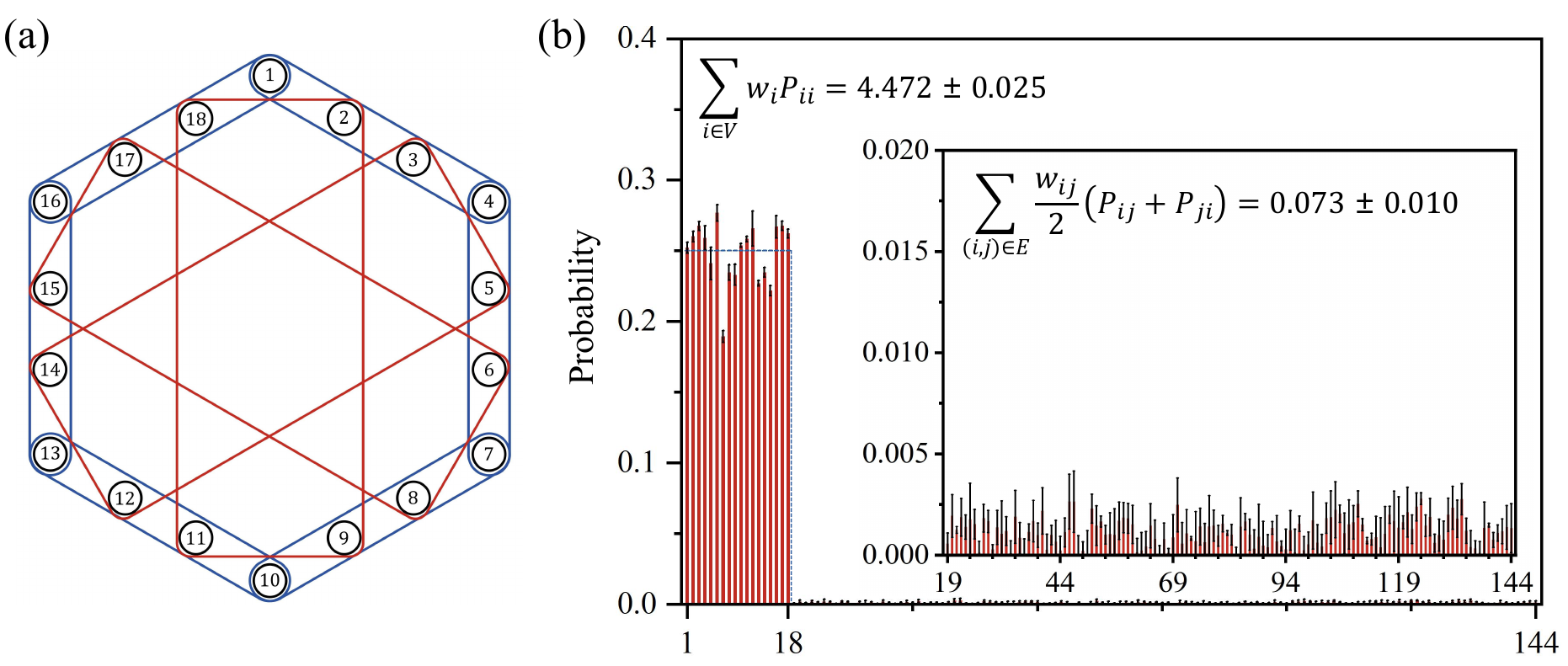}
\caption{\sf (a) Graph of orthogonality between the vectors of the KS18 set \cite{cabello1996bell}, the vertice with number \(i\) represents each vector \(v_i\) of KS18, \( i\in \left \{ 1,\dots ,18 \right \}\). Each of the 6 sides of the regular hexagon (blue) and each of the 3 rectangles (red) represent the contexts that contain only orthogonal vectors, and all of the vertices are weighted 1. (b) Detailed experimental results leading to \(\beta\left(\text {KS18}\right)=4.399 \pm 0.027\).}
\label{fig3}
\end{figure*}

\begin{figure*}[!tb]
\centering
\includegraphics[width=15cm]{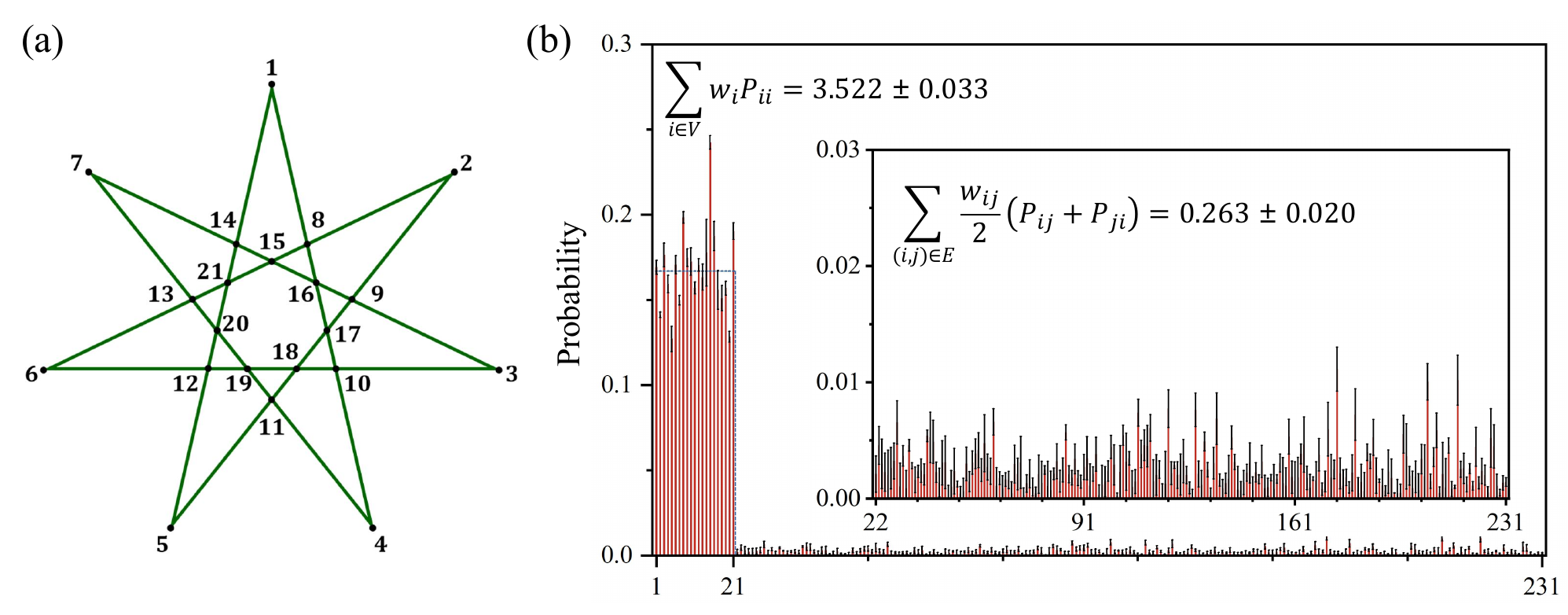}
\caption{\sf (a) Graph of orthogonality between the vectors of the KS21 set \cite{lisonvek2014kochen}, the vertice with number \(i\) represents each vector \(v_i\) of KS21, \( i\in \left \{ 1,\dots ,21 \right \}\). Each straight line in the great heptagram represents the contexts that contain only orthogonal vectors, and all of the vertices are weighted 1. (b) Detailed experimental results leading to \(\beta\left(\text {KS21}\right)=3.259 \pm 0.038\).}
\label{fig4}
\end{figure*}

SI-C is a specific contextuality argument that depends not on the choice of a particular quantum state, which is directly related to proofs of the Kochen-Specker (KS) theorem \cite{kochen1990problem}. 
A SI-C set is critical if by removing any of its elements the resulting set is not a SI-C set, and a minimal critical SI-C set is a critical SI-C set of minimum cardinality. We consider several critical SI-C sets for which minimality has been proven and give the orthogonality graph, respectively. The simplest SI-C set is the one with 13 projectors in dimension \(d=3\), theoretically found by Yu and Oh \cite{Yu2011}, here donated as YO13, and represented the orthogonality graph in FIG. \ref{fig2}(a). The simplest KS set in dimension \(d=4\) is the 18-vector (nine-context) set proposed by Cabello and coworkers \cite{cabello1996bell}, here called KS18 and shown in FIG. \ref{fig3}(a). The KS set with the smallest number of contexts known is the 21-vector, seven-context set in dimension \(d=6\) introduced by Lison{\v{e}}k \textit{et al.} \cite{lisonvek2014kochen}, here called KS21 and shown in FIG. \ref{fig4}(a).

Following the method introduced in \cite{cabello2021converting}, one can construct a Bell inequality by embedding a SI-C set into the bipartite scenario \cite{stairs1983quantum,heywood1983nonlocality}. However, experimental tests of the inequalities are challenging and noise-sensitive, and the requirements for the quantum states purity, dimensionality, and degree of entanglement have blocked the experimental implementation. Distinguishing from the polarization, which is naturally bidimensional, orbital angular momentum (OAM) is infinite-dimensional, such as a photon with \(\ell \) intertwined helical wavefronts, \(\left | \ell  \right \rangle\), carries \(\ell\hbar\) units of OAM \cite{mair2001entanglement}. As these modes can be used to define an infinitely dimensional discrete Hilbert space, and the number of effective dimensions can be readily tailored as required, this approach provides a practical route to entanglement with higher dimensions \cite{torres2003quantum}. The spontaneous parametric down-conversion (SPDC) process conserves OAM, \(\ell \) can take on any integer value that is theoretically infinite-dimensional, although in practice the apertures in the experiment limit the dimensions \cite{leach2010quantum}. Controlling OAM state superpositions opens the door to generating and manipulating multi-dimensional quantum states \cite{erhard2018twisted,padgett2004light,molina2001management}. By exploiting two-photon OAM high-dimensional entangled states, we experimentally test the Bell inequalities from SI-C sets based on the theoretical work of Cabello \cite{cabello2021converting}. 

Our experimental setup is shown in FIG. \ref{fig1}. We start by generating an entangled pair of photons through the SPDC process, by sending a mode-locked 355 nm Nd-YAG laser through a BBO crystal. We use a long-pass filter behind the crystal to block the pump beam, and then a 50:50 beam splitter (BS) separates the signal and idler photons. The conservation of orbital angular momentum ensures that if the signal photon is in the mode specified by $\left | \ell  \right \rangle$, the corresponding idler photon can only be in the mode $\left | -\ell  \right \rangle$~\cite{walborn2004entanglement}. Through this process, we generate photon pairs entangled according to the non-separable wavefunction \(\left | \Psi  \right \rangle =\sum_{\ell=-\infty }^{\ell=\infty}c_{\ell} \left | \ell  \right \rangle_{A}\otimes\left | -\ell  \right \rangle_{B}\), where \(c_{\ell}\) is the probability amplitude that the photon in arm \(A\) is in the state \(\left | \ell  \right \rangle_{A}\) and the photon in arm \(B\) is in \(\left | -\ell  \right \rangle_{B}\). In the detection, we use computer-controlled spatial light modulators (SLM) operating in reflection mode loaded with specially designed holographic gratings, both for preparing the desired measurement OAM states (see Supplementary Material \cite{supp} for more details of the holograms corresponding to measurement vectors). An SLM prepared in a given state transforms a photon in that state to the Gaussian \(|\ell=0\rangle\) mode. The hologram generation algorithm introduced in Ref. \cite{leach2005vortex} is applied to configure the SLMs. Then a 4-\textit{f} telescope consisting of two lenses is used to reimage the reflected photon and coupled into a single-mode fiber (SMF) which feeds an avalanche photon detector. Since only the \(|\ell=0\rangle\) mode is coupled into the fiber, the count in the detector indicates detection of the state in which the SLM was prepared. Band-pass filters (BF) with 10 nm width are placed in front of the SMF to reduce the detection of noise photons. The outputs of the detectors are connected to a coincidence circuit, the coincidence resolving time is 10 ns. 

The spiral bandwidth is an important factor affecting the degree of entanglement of the photon pairs for a given selected subspace of the OAM Hilbert space \cite{torres2003quantum}. As shown in FIG. \ref{fig1}(a), the finiteness of the spiral bandwidth leads to a non-maximally entangled state for the projection of the SPDC output state onto a \textit{d}-dimensional subspace. We perform the Procrustean filtering technique for entanglement concentration to enhance the entanglement \cite{bennett1996concentrating,law2004analysis}, which can be considered to be generalized measurements performed by local operations on the signal and idler beams \cite{supp}. By exploiting alterations in the diffraction efficiencies of blazed phase gratings in the SLMs, we obtain a close approximation to a maximally entangled state \(\left|\psi_d\right\rangle=\frac{1}{\sqrt{d}} \sum_{j=0}^{d-1}|j\rangle_A|j\rangle_B\), as shown in FIG. \ref{fig1}(b). The disadvantage of Procrustean filtering is the reduction in photon detection efficiency~\cite{dada2011experimental}. Due to the experimental imperfections, such as optical misalignment, we adopt suitable OAM intervals $\Delta \ell$ to perform the measurements. Such a trade-off consideration enables us to obtain a relatively low crosstalk between orthogonal modes yet a relatively high coincidence count. For the tests within the subspaces of \(d=3,4\), we prioritized higher extinction between the orthogonal states, while compensating for lower rates with longer integration times. We choose the modes \(\ell=-3, 0, 3\) as the computational basis states for \(d=3\), and  \(\ell=-4, -1,1, 4\) for \(d=4\). For the test within the subspace of \(d=6\), we choose the modes \(\ell=-3,-2,-1,1,2,3\) as the computational basis, for a higher count rate within fewer integration times. 

Given a SI-C set \(\left\{\Pi_i\right\}\), where \(\Pi_i=\left|v_i\right\rangle\left\langle v_i\right|\), we consider the inequality from the set
\begin{align}\label{Bellinequality}
\sum_{i\in V(\mathcal{G})} w_i P_{ii}-\sum_{(i, j)\in E(\mathcal{G})} \frac{w_{ij}}{2} \left ( P_{ij}+P_{ji} \right )\leq \alpha(\mathcal{G}, w) ,
\end{align}
where \(P_{ij}:=P\left(\Pi_i^A=1, \Pi_j^B=1\right)\) is the probability that Alice obtains outcome \(1\) for measurement \(\Pi_i\) on her particle and Bob obtains outcome \(1\) for measurement \(\Pi_j\) on his particle. \(\mathcal{G}\) is the compatibility graph associated with the SI-C, \(V(\mathcal{G})\) and \(E(\mathcal{G})\) are the sets of vertices and edges of \(\mathcal{G}\), respectively, \(\alpha(\mathcal{G}, w)\) is the independence number of the weighted graph \((\mathcal{G}, w)\). For convenience, we denote that \(w_{ij}:=\max \left(w_i, w_j\right)\) and the left-hand side of inequality \eqref{Bellinequality} as \(\beta(\mathcal{G})\) in the following. 

Experimentally, we test the inequality \eqref{Bellinequality} from the YO13 set \(\left\{\Pi_i\right\}_{i=1}^{13}\). Distributing pairs of particles in the 3-dimensional OAM maximally entangled state, \(\left|\psi_3\right\rangle=\frac{1}{\sqrt{3}} \sum_{j=0}^{2}|j\rangle_A|j\rangle_B\), between Alice and Bob and allowing each of them to perform a randomly chosen spacelike separated measurement from the YO13 set. The weights that lead to the largest gap between the quantum violation and the classical bound is \(w_i=3\) for all projectors, except for those onto the vertices in red (\(v_3\), \(v_7\), \(v_{12}\), \(v_{13}\)), where \(w_i=2\). With these weights, the classical bound of inequality \eqref{Bellinequality} is given by the independence number of the graph \(\alpha(\text {YO13})=11\) \cite{cabello2014graph}. Theoretically, the quantum maximum observation of \(\beta\left(\text {YO13}\right)\) cannot be higher than \(\vartheta\left(\text {YO13}\right)=11+\frac{2}{3}\), which is given by the weighted Lov{\'{a}}sz number \cite{lovasz1979shannon}. We experimentally observed that 
\begin{align}
 \beta\left(\text {YO13}\right)=11.573 \pm 0.012,
\end{align}
which violates the inequality \eqref{Bellinequality} by 48 standard deviations. Experimental results of the correlations in the inequality \eqref{Bellinequality} are shown in FIG. \ref{fig2}(b). For simplicity, in FIG. \ref{fig2}(b), we use the numbers 1 to 13 to represent the joint probability corresponding to each of the vertices \(v_i\), and 14 to 61 to denote the joint probability corresponding to the 48 edges of the vertex-weighted graph of the YO13 set, respectively, as well as within the context sets following. More detailed experimental results can be found in the Supplemental Material \cite{[{See Supplemental Material at }][{, which includes Refs. [36–41], for more details on the derivation of inequality (1), entanglement concentration, experimental data, and discussion of loopholes.}]supp}.

Similarly, in the 4-dimensional OAM maximally entangled state, \(\left|\psi_4\right\rangle=\frac{1}{2} \sum_{j=0}^{3}|j\rangle_A|j\rangle_B\), we perform randomly chosen local measurements on each party from the KS18 set \(\left\{\Pi_i\right\}_{i=1}^{18}\). The weights that lead to the largest gap between the quantum violation and the classical bound are \(\left\{w_i=1\right\}_{i=1}^{18}\). With these weights, the upper bound is \(\alpha(\text {KS18})=4\), and the maximum observation allowed by quantum mechanics is \(\vartheta(\text {KS18})=4.5\). As shown in FIG. \ref{fig3}(b), the experimental data lead to that 
\begin{align}
 \beta\left(\text {KS18}\right)=4.399 \pm 0.027,
\end{align}
which violates the inequality \eqref{Bellinequality} by 15 standard deviations.

The KS21 set is the SI-C set with the smallest number of contexts (seven-context) allowing for a symmetric parity proof of the KS theorem.  In the 6-dimensional OAM maximally entangled state, \(\left|\psi_6\right\rangle=\frac{1}{\sqrt{6}} \sum_{j=0}^{5}|j\rangle_A|j\rangle_B\), each separated photon performed a randomly chosen measurement from the KS21 set \(\left\{\Pi_i\right\}_{i=1}^{21}\). The weights that lead to the largest gap between the quantum violation and the classical bound are \(\left\{w_i=1\right\}_{i=1}^{21}\). With these weights, \(\alpha(\text {KS21})=3\), and \(\vartheta(\text {KS21})=3.5\). We experimentally obtained that 
\begin{align}
 \beta\left(\text {KS21}\right)=3.259 \pm 0.038,
\end{align}
which violates the inequality \eqref{Bellinequality} by 7 standard deviations. All errors were calculated assuming Poisson statistics for the photon counting processes. 
In the Supplementary Material, we also employed another quantitative metric, the \textit{p}-value analysis, to help with the error analysis. The results indicate that there is a statistically significant difference between the experimental observation and the classical bound, which suggests that the violation of inequalities is not due to random chance \cite{supp}.

In conclusion, we experimentally test the Bell inequalities related to the SI-C sets \cite{supp}. 
Our results show that some correlations in bipartite systems cannot be reproduced with noncontextual hidden variable theories. 
In previous experimental tests of contextuality, performing the joint measurements in single particle systems implies more complex experimental setups, and the need to perform additional experimental runs to quantify deviations from ideal measurements \cite{kirchmair2009state,guhne2010compatibility,amselem2009state,lapkiewicz2011experimental}. 
Distinguishing from the standard single-particle ways, our experiment spotlights that the SI-C sets can also be tested in a bipartite scenario, by which ``compatibility'' \cite{szangolies2013tests,szangolies2015testing} or ``sharpness'' \cite{vallee2024corrected,spekkens2014status,kunjwal2015kochen,krishna2017deriving} loopholes can be effectively avoided, 
and opening up the possibility of loophole-free contextuality testing \cite{wang2022significant,hu2023self} on optical systems (see also \cite{supp} for more discussion). 
For the high-dimensional SI-C sets that are challenging to test on single-particle systems, our work can be of some experimental indicativeness. 
Furthermore, the quantum violations of the SI-C inequality and the Bell inequality can be tested simultaneously in the same experiment, by allowing one of the parties to perform sequential measurements. According to quantum theory, the experiment would give equal violations of both inequalities, in which some interesting aspects remain to be further explored in future studies.

\begin{acknowledgments} 
This work was supported by National Natural Science Foundation of China (12034016, 12205107), the National Key R\&D Program of China (2023YFA1407200), Natural Science Foundation of Fujian Province of China (2021J02002) for Distinguished Young Scientists (2015J06002), Program for New Century Excellent Talents in University (NCET-13-0495), and Natural Science Foundation of Xiamen City (3502Z20227033).
\end{acknowledgments}

\bibliographystyle{apsrev4-2} 

\begin{thebibliography}{54}%
	\makeatletter
	\providecommand \@ifxundefined [1]{%
		\@ifx{#1\undefined}
	}%
	\providecommand \@ifnum [1]{%
		\ifnum #1\expandafter \@firstoftwo
		\else \expandafter \@secondoftwo
		\fi
	}%
	\providecommand \@ifx [1]{%
		\ifx #1\expandafter \@firstoftwo
		\else \expandafter \@secondoftwo
		\fi
	}%
	\providecommand \natexlab [1]{#1}%
	\providecommand \enquote  [1]{``#1''}%
	\providecommand \bibnamefont  [1]{#1}%
	\providecommand \bibfnamefont [1]{#1}%
	\providecommand \citenamefont [1]{#1}%
	\providecommand \href@noop [0]{\@secondoftwo}%
	\providecommand \href [0]{\begingroup \@sanitize@url \@href}%
	\providecommand \@href[1]{\@@startlink{#1}\@@href}%
	\providecommand \@@href[1]{\endgroup#1\@@endlink}%
	\providecommand \@sanitize@url [0]{\catcode `\\12\catcode `\$12\catcode
		`\&12\catcode `\#12\catcode `\^12\catcode `\_12\catcode `\%12\relax}%
	\providecommand \@@startlink[1]{}%
	\providecommand \@@endlink[0]{}%
	\providecommand \url  [0]{\begingroup\@sanitize@url \@url }%
	\providecommand \@url [1]{\endgroup\@href {#1}{\urlprefix }}%
	\providecommand \urlprefix  [0]{URL }%
	\providecommand \Eprint [0]{\href }%
	\providecommand \doibase [0]{https://doi.org/}%
	\providecommand \selectlanguage [0]{\@gobble}%
	\providecommand \bibinfo  [0]{\@secondoftwo}%
	\providecommand \bibfield  [0]{\@secondoftwo}%
	\providecommand \translation [1]{[#1]}%
	\providecommand \BibitemOpen [0]{}%
	\providecommand \bibitemStop [0]{}%
	\providecommand \bibitemNoStop [0]{.\EOS\space}%
	\providecommand \EOS [0]{\spacefactor3000\relax}%
	\providecommand \BibitemShut  [1]{\csname bibitem#1\endcsname}%
	\let\auto@bib@innerbib\@empty
	\bibitem [{\citenamefont {Brunner}\ \emph {et~al.}(2014)\citenamefont
		{Brunner}, \citenamefont {Cavalcanti}, \citenamefont {Pironio}, \citenamefont
		{Scarani},\ and\ \citenamefont {Wehner}}]{brunner2014bell}%
	\BibitemOpen
	\bibfield  {author} {\bibinfo {author} {\bibfnamefont {N.}~\bibnamefont
			{Brunner}}, \bibinfo {author} {\bibfnamefont {D.}~\bibnamefont {Cavalcanti}},
		\bibinfo {author} {\bibfnamefont {S.}~\bibnamefont {Pironio}}, \bibinfo
		{author} {\bibfnamefont {V.}~\bibnamefont {Scarani}},\ and\ \bibinfo {author}
		{\bibfnamefont {S.}~\bibnamefont {Wehner}},\ }\href@noop {} {\bibfield
		{journal} {\bibinfo  {journal} {Reviews of modern physics}\ }\textbf
		{\bibinfo {volume} {86}},\ \bibinfo {pages} {419} (\bibinfo {year}
		{2014})}\BibitemShut {NoStop}%
	\bibitem [{\citenamefont {Pironio}\ \emph {et~al.}(2010)\citenamefont
		{Pironio}, \citenamefont {Ac{\'\i}n}, \citenamefont {Massar}, \citenamefont
		{de~La~Giroday}, \citenamefont {Matsukevich}, \citenamefont {Maunz},
		\citenamefont {Olmschenk}, \citenamefont {Hayes}, \citenamefont {Luo},
		\citenamefont {Manning} \emph {et~al.}}]{pironio2010random}%
	\BibitemOpen
	\bibfield  {author} {\bibinfo {author} {\bibfnamefont {S.}~\bibnamefont
			{Pironio}}, \bibinfo {author} {\bibfnamefont {A.}~\bibnamefont {Ac{\'\i}n}},
		\bibinfo {author} {\bibfnamefont {S.}~\bibnamefont {Massar}}, \bibinfo
		{author} {\bibfnamefont {A.~B.}\ \bibnamefont {de~La~Giroday}}, \bibinfo
		{author} {\bibfnamefont {D.~N.}\ \bibnamefont {Matsukevich}}, \bibinfo
		{author} {\bibfnamefont {P.}~\bibnamefont {Maunz}}, \bibinfo {author}
		{\bibfnamefont {S.}~\bibnamefont {Olmschenk}}, \bibinfo {author}
		{\bibfnamefont {D.}~\bibnamefont {Hayes}}, \bibinfo {author} {\bibfnamefont
			{L.}~\bibnamefont {Luo}}, \bibinfo {author} {\bibfnamefont {T.~A.}\
			\bibnamefont {Manning}}, \emph {et~al.},\ }\href@noop {} {\bibfield
		{journal} {\bibinfo  {journal} {Nature}\ }\textbf {\bibinfo {volume} {464}},\
		\bibinfo {pages} {1021} (\bibinfo {year} {2010})}\BibitemShut {NoStop}%
	\bibitem [{\citenamefont {Ac{\'\i}n}\ \emph {et~al.}(2007)\citenamefont
		{Ac{\'\i}n}, \citenamefont {Brunner}, \citenamefont {Gisin}, \citenamefont
		{Massar}, \citenamefont {Pironio},\ and\ \citenamefont
		{Scarani}}]{acin2007device}%
	\BibitemOpen
	\bibfield  {author} {\bibinfo {author} {\bibfnamefont {A.}~\bibnamefont
			{Ac{\'\i}n}}, \bibinfo {author} {\bibfnamefont {N.}~\bibnamefont {Brunner}},
		\bibinfo {author} {\bibfnamefont {N.}~\bibnamefont {Gisin}}, \bibinfo
		{author} {\bibfnamefont {S.}~\bibnamefont {Massar}}, \bibinfo {author}
		{\bibfnamefont {S.}~\bibnamefont {Pironio}},\ and\ \bibinfo {author}
		{\bibfnamefont {V.}~\bibnamefont {Scarani}},\ }\href@noop {} {\bibfield
		{journal} {\bibinfo  {journal} {Physical Review Letters}\ }\textbf {\bibinfo
			{volume} {98}},\ \bibinfo {pages} {230501} (\bibinfo {year}
		{2007})}\BibitemShut {NoStop}%
	\bibitem [{\citenamefont {Budroni}\ \emph {et~al.}(2022)\citenamefont
		{Budroni}, \citenamefont {Cabello}, \citenamefont {G{\"u}hne}, \citenamefont
		{Kleinmann},\ and\ \citenamefont {Larsson}}]{budroni2022kochen}%
	\BibitemOpen
	\bibfield  {author} {\bibinfo {author} {\bibfnamefont {C.}~\bibnamefont
			{Budroni}}, \bibinfo {author} {\bibfnamefont {A.}~\bibnamefont {Cabello}},
		\bibinfo {author} {\bibfnamefont {O.}~\bibnamefont {G{\"u}hne}}, \bibinfo
		{author} {\bibfnamefont {M.}~\bibnamefont {Kleinmann}},\ and\ \bibinfo
		{author} {\bibfnamefont {J.-{\AA}.}\ \bibnamefont {Larsson}},\ }\href@noop {}
	{\bibfield  {journal} {\bibinfo  {journal} {Reviews of Modern Physics}\
		}\textbf {\bibinfo {volume} {94}},\ \bibinfo {pages} {045007} (\bibinfo
		{year} {2022})}\BibitemShut {NoStop}%
	\bibitem [{\citenamefont {Cubitt}\ \emph {et~al.}(2010)\citenamefont {Cubitt},
		\citenamefont {Leung}, \citenamefont {Matthews},\ and\ \citenamefont
		{Winter}}]{cubitt2010improving}%
	\BibitemOpen
	\bibfield  {author} {\bibinfo {author} {\bibfnamefont {T.~S.}\ \bibnamefont
			{Cubitt}}, \bibinfo {author} {\bibfnamefont {D.}~\bibnamefont {Leung}},
		\bibinfo {author} {\bibfnamefont {W.}~\bibnamefont {Matthews}},\ and\
		\bibinfo {author} {\bibfnamefont {A.}~\bibnamefont {Winter}},\ }\href@noop {}
	{\bibfield  {journal} {\bibinfo  {journal} {Physical Review Letters}\
		}\textbf {\bibinfo {volume} {104}},\ \bibinfo {pages} {230503} (\bibinfo
		{year} {2010})}\BibitemShut {NoStop}%
	\bibitem [{\citenamefont {Saha}\ \emph {et~al.}(2019)\citenamefont {Saha},
		\citenamefont {Horodecki},\ and\ \citenamefont
		{Paw{\l}owski}}]{saha2019state}%
	\BibitemOpen
	\bibfield  {author} {\bibinfo {author} {\bibfnamefont {D.}~\bibnamefont
			{Saha}}, \bibinfo {author} {\bibfnamefont {P.}~\bibnamefont {Horodecki}},\
		and\ \bibinfo {author} {\bibfnamefont {M.}~\bibnamefont {Paw{\l}owski}},\
	}\href@noop {} {\bibfield  {journal} {\bibinfo  {journal} {New Journal of
				Physics}\ }\textbf {\bibinfo {volume} {21}},\ \bibinfo {pages} {093057}
		(\bibinfo {year} {2019})}\BibitemShut {NoStop}%
	\bibitem [{\citenamefont {Gupta}\ \emph {et~al.}(2023)\citenamefont {Gupta},
		\citenamefont {Saha}, \citenamefont {Xu}, \citenamefont {Cabello},\ and\
		\citenamefont {Majumdar}}]{gupta2023quantum}%
	\BibitemOpen
	\bibfield  {author} {\bibinfo {author} {\bibfnamefont {S.}~\bibnamefont
			{Gupta}}, \bibinfo {author} {\bibfnamefont {D.}~\bibnamefont {Saha}},
		\bibinfo {author} {\bibfnamefont {Z.-P.}\ \bibnamefont {Xu}}, \bibinfo
		{author} {\bibfnamefont {A.}~\bibnamefont {Cabello}},\ and\ \bibinfo {author}
		{\bibfnamefont {A.~S.}\ \bibnamefont {Majumdar}},\ }\href@noop {} {\bibfield
		{journal} {\bibinfo  {journal} {Physical Review Letters}\ }\textbf {\bibinfo
			{volume} {130}},\ \bibinfo {pages} {080802} (\bibinfo {year}
		{2023})}\BibitemShut {NoStop}%
	\bibitem [{\citenamefont {Cabello}\ \emph {et~al.}(2011)\citenamefont
		{Cabello}, \citenamefont {D鈥橝mbrosio}, \citenamefont {Nagali},\ and\
		\citenamefont {Sciarrino}}]{cabello2011hybrid}%
	\BibitemOpen
	\bibfield  {author} {\bibinfo {author} {\bibfnamefont {A.}~\bibnamefont
			{Cabello}}, \bibinfo {author} {\bibfnamefont {V.}~\bibnamefont
			{D'mbrosio}}, \bibinfo {author} {\bibfnamefont {E.}~\bibnamefont
			{Nagali}},\ and\ \bibinfo {author} {\bibfnamefont {F.}~\bibnamefont
			{Sciarrino}},\ }\href@noop {} {\bibfield  {journal} {\bibinfo  {journal}
			{Physical Review A-Atomic, Molecular, and Optical Physics}\ }\textbf
		{\bibinfo {volume} {84}},\ \bibinfo {pages} {030302} (\bibinfo {year}
		{2011})}\BibitemShut {NoStop}%
	\bibitem [{\citenamefont {Zhen}\ \emph {et~al.}(2023)\citenamefont {Zhen},
		\citenamefont {Mao}, \citenamefont {Zhang}, \citenamefont {Xu},\ and\
		\citenamefont {Sanders}}]{zhen2023device}%
	\BibitemOpen
	\bibfield  {author} {\bibinfo {author} {\bibfnamefont {Y.-Z.}\ \bibnamefont
			{Zhen}}, \bibinfo {author} {\bibfnamefont {Y.}~\bibnamefont {Mao}}, \bibinfo
		{author} {\bibfnamefont {Y.-Z.}\ \bibnamefont {Zhang}}, \bibinfo {author}
		{\bibfnamefont {F.}~\bibnamefont {Xu}},\ and\ \bibinfo {author}
		{\bibfnamefont {B.~C.}\ \bibnamefont {Sanders}},\ }\href@noop {} {\bibfield
		{journal} {\bibinfo  {journal} {Physical Review Letters}\ }\textbf {\bibinfo
			{volume} {131}},\ \bibinfo {pages} {080801} (\bibinfo {year}
		{2023})}\BibitemShut {NoStop}%
	\bibitem [{\citenamefont {Raussendorf}(2013)}]{raussendorf2013contextuality}%
	\BibitemOpen
	\bibfield  {author} {\bibinfo {author} {\bibfnamefont {R.}~\bibnamefont
			{Raussendorf}},\ }\href {https://doi.org/10.1103/PhysRevA.88.022322}
	{\bibfield  {journal} {\bibinfo  {journal} {Phys. Rev. A}\ }\textbf {\bibinfo
			{volume} {88}},\ \bibinfo {pages} {022322} (\bibinfo {year}
		{2013})}\BibitemShut {NoStop}%
	\bibitem [{\citenamefont {Delfosse}\ \emph {et~al.}(2015)\citenamefont
		{Delfosse}, \citenamefont {Allard~Guerin}, \citenamefont {Bian},\ and\
		\citenamefont {Raussendorf}}]{delfosse2015wigner}%
	\BibitemOpen
	\bibfield  {author} {\bibinfo {author} {\bibfnamefont {N.}~\bibnamefont
			{Delfosse}}, \bibinfo {author} {\bibfnamefont {P.}~\bibnamefont
			{Allard~Guerin}}, \bibinfo {author} {\bibfnamefont {J.}~\bibnamefont
			{Bian}},\ and\ \bibinfo {author} {\bibfnamefont {R.}~\bibnamefont
			{Raussendorf}},\ }\href@noop {} {\bibfield  {journal} {\bibinfo  {journal}
			{Physical Review X}\ }\textbf {\bibinfo {volume} {5}},\ \bibinfo {pages}
		{021003} (\bibinfo {year} {2015})}\BibitemShut {NoStop}%
	\bibitem [{\citenamefont {Bravyi}\ \emph {et~al.}(2018)\citenamefont {Bravyi},
		\citenamefont {Gosset},\ and\ \citenamefont {K{\"o}nig}}]{bravyi2018quantum}%
	\BibitemOpen
	\bibfield  {author} {\bibinfo {author} {\bibfnamefont {S.}~\bibnamefont
			{Bravyi}}, \bibinfo {author} {\bibfnamefont {D.}~\bibnamefont {Gosset}},\
		and\ \bibinfo {author} {\bibfnamefont {R.}~\bibnamefont {K{\"o}nig}},\
	}\href@noop {} {\bibfield  {journal} {\bibinfo  {journal} {Science}\ }\textbf
		{\bibinfo {volume} {362}},\ \bibinfo {pages} {308} (\bibinfo {year}
		{2018})}\BibitemShut {NoStop}%
	\bibitem [{\citenamefont {Stairs}(1983)}]{stairs1983quantum}%
	\BibitemOpen
	\bibfield  {author} {\bibinfo {author} {\bibfnamefont {A.}~\bibnamefont
			{Stairs}},\ }\href@noop {} {\bibfield  {journal} {\bibinfo  {journal}
			{Philosophy of Science}\ }\textbf {\bibinfo {volume} {50}},\ \bibinfo {pages}
		{578} (\bibinfo {year} {1983})}\BibitemShut {NoStop}%
	\bibitem [{\citenamefont {Heywood}\ and\ \citenamefont
		{Redhead}(1983)}]{heywood1983nonlocality}%
	\BibitemOpen
	\bibfield  {author} {\bibinfo {author} {\bibfnamefont {P.}~\bibnamefont
			{Heywood}}\ and\ \bibinfo {author} {\bibfnamefont {M.~L.}\ \bibnamefont
			{Redhead}},\ }\href@noop {} {\bibfield  {journal} {\bibinfo  {journal}
			{Foundations of physics}\ }\textbf {\bibinfo {volume} {13}},\ \bibinfo
		{pages} {481} (\bibinfo {year} {1983})}\BibitemShut {NoStop}%
	\bibitem [{\citenamefont {Abramsky}\ and\ \citenamefont
		{Brandenburger}(2011)}]{abramsky2011sheaf}%
	\BibitemOpen
	\bibfield  {author} {\bibinfo {author} {\bibfnamefont {S.}~\bibnamefont
			{Abramsky}}\ and\ \bibinfo {author} {\bibfnamefont {A.}~\bibnamefont
			{Brandenburger}},\ }\href@noop {} {\bibfield  {journal} {\bibinfo  {journal}
			{New Journal of Physics}\ }\textbf {\bibinfo {volume} {13}},\ \bibinfo
		{pages} {113036} (\bibinfo {year} {2011})}\BibitemShut {NoStop}%
	\bibitem [{\citenamefont {Abramsky}\ \emph {et~al.}(2011)\citenamefont
		{Abramsky}, \citenamefont {Mansfield},\ and\ \citenamefont
		{Barbosa}}]{abramsky2011cohomology}%
	\BibitemOpen
	\bibfield  {author} {\bibinfo {author} {\bibfnamefont {S.}~\bibnamefont
			{Abramsky}}, \bibinfo {author} {\bibfnamefont {S.}~\bibnamefont
			{Mansfield}},\ and\ \bibinfo {author} {\bibfnamefont {R.~S.}\ \bibnamefont
			{Barbosa}},\ }\href@noop {} {\bibfield  {journal} {\bibinfo  {journal} {arXiv
				preprint arXiv:1111.3620}\ } (\bibinfo {year} {2011})}\BibitemShut {NoStop}%
	\bibitem [{\citenamefont {Amaral}\ \emph {et~al.}(2018)\citenamefont {Amaral},
		\citenamefont {Cabello}, \citenamefont {Cunha},\ and\ \citenamefont
		{Aolita}}]{amaral2018noncontextual}%
	\BibitemOpen
	\bibfield  {author} {\bibinfo {author} {\bibfnamefont {B.}~\bibnamefont
			{Amaral}}, \bibinfo {author} {\bibfnamefont {A.}~\bibnamefont {Cabello}},
		\bibinfo {author} {\bibfnamefont {M.~T.}\ \bibnamefont {Cunha}},\ and\
		\bibinfo {author} {\bibfnamefont {L.}~\bibnamefont {Aolita}},\ }\href@noop {}
	{\bibfield  {journal} {\bibinfo  {journal} {Physical review letters}\
		}\textbf {\bibinfo {volume} {120}},\ \bibinfo {pages} {130403} (\bibinfo
		{year} {2018})}\BibitemShut {NoStop}%
	\bibitem [{\citenamefont {Abramsky}\ and\ \citenamefont
		{Barbosa}(2020)}]{abramsky2020logic}%
	\BibitemOpen
	\bibfield  {author} {\bibinfo {author} {\bibfnamefont {S.}~\bibnamefont
			{Abramsky}}\ and\ \bibinfo {author} {\bibfnamefont {R.~S.}\ \bibnamefont
			{Barbosa}},\ }\href@noop {} {\bibfield  {journal} {\bibinfo  {journal} {arXiv
				preprint arXiv:2011.03064}\ } (\bibinfo {year} {2020})}\BibitemShut {NoStop}%
	\bibitem [{\citenamefont {Cabello}(2001)}]{cabello2001all}%
	\BibitemOpen
	\bibfield  {author} {\bibinfo {author} {\bibfnamefont {A.}~\bibnamefont
			{Cabello}},\ }\href@noop {} {\bibfield  {journal} {\bibinfo  {journal}
			{Physical Review Letters}\ }\textbf {\bibinfo {volume} {87}},\ \bibinfo
		{pages} {010403} (\bibinfo {year} {2001})}\BibitemShut {NoStop}%
	\bibitem [{\citenamefont {Aolita}\ \emph {et~al.}(2012)\citenamefont {Aolita},
		\citenamefont {Gallego}, \citenamefont {Ac{\'\i}n}, \citenamefont {Chiuri},
		\citenamefont {Vallone}, \citenamefont {Mataloni},\ and\ \citenamefont
		{Cabello}}]{aolita2012fully}%
	\BibitemOpen
	\bibfield  {author} {\bibinfo {author} {\bibfnamefont {L.}~\bibnamefont
			{Aolita}}, \bibinfo {author} {\bibfnamefont {R.}~\bibnamefont {Gallego}},
		\bibinfo {author} {\bibfnamefont {A.}~\bibnamefont {Ac{\'\i}n}}, \bibinfo
		{author} {\bibfnamefont {A.}~\bibnamefont {Chiuri}}, \bibinfo {author}
		{\bibfnamefont {G.}~\bibnamefont {Vallone}}, \bibinfo {author} {\bibfnamefont
			{P.}~\bibnamefont {Mataloni}},\ and\ \bibinfo {author} {\bibfnamefont
			{A.}~\bibnamefont {Cabello}},\ }\href@noop {} {\bibfield  {journal} {\bibinfo
			{journal} {Physical Review A-Atomic, Molecular, and Optical Physics}\
		}\textbf {\bibinfo {volume} {85}},\ \bibinfo {pages} {032107} (\bibinfo
		{year} {2012})}\BibitemShut {NoStop}%
	\bibitem [{\citenamefont {Wright}\ and\ \citenamefont
		{Kunjwal}(2023)}]{wright2023contextuality}%
	\BibitemOpen
	\bibfield  {author} {\bibinfo {author} {\bibfnamefont {V.~J.}\ \bibnamefont
			{Wright}}\ and\ \bibinfo {author} {\bibfnamefont {R.}~\bibnamefont
			{Kunjwal}},\ }\href@noop {} {\bibfield  {journal} {\bibinfo  {journal}
			{Quantum}\ }\textbf {\bibinfo {volume} {7}},\ \bibinfo {pages} {900}
		(\bibinfo {year} {2023})}\BibitemShut {NoStop}%
	\bibitem [{\citenamefont {Liu}\ \emph {et~al.}(2016)\citenamefont {Liu},
		\citenamefont {Hu}, \citenamefont {Chen}, \citenamefont {Huang},
		\citenamefont {Han}, \citenamefont {Li}, \citenamefont {Guo},\ and\
		\citenamefont {Cabello}}]{liu2016nonlocality}%
	\BibitemOpen
	\bibfield  {author} {\bibinfo {author} {\bibfnamefont {B.-H.}\ \bibnamefont
			{Liu}}, \bibinfo {author} {\bibfnamefont {X.-M.}\ \bibnamefont {Hu}},
		\bibinfo {author} {\bibfnamefont {J.-S.}\ \bibnamefont {Chen}}, \bibinfo
		{author} {\bibfnamefont {Y.-F.}\ \bibnamefont {Huang}}, \bibinfo {author}
		{\bibfnamefont {Y.-J.}\ \bibnamefont {Han}}, \bibinfo {author} {\bibfnamefont
			{C.-F.}\ \bibnamefont {Li}}, \bibinfo {author} {\bibfnamefont {G.-C.}\
			\bibnamefont {Guo}},\ and\ \bibinfo {author} {\bibfnamefont {A.}~\bibnamefont
			{Cabello}},\ }\href@noop {} {\bibfield  {journal} {\bibinfo  {journal}
			{Physical Review Letters}\ }\textbf {\bibinfo {volume} {117}},\ \bibinfo
		{pages} {220402} (\bibinfo {year} {2016})}\BibitemShut {NoStop}%
	\bibitem [{\citenamefont {Xue}\ \emph {et~al.}(2023)\citenamefont {Xue},
		\citenamefont {Xiao}, \citenamefont {Ruffolo}, \citenamefont {Mazzari},
		\citenamefont {Temistocles}, \citenamefont {Cunha},\ and\ \citenamefont
		{Rabelo}}]{xue2023synchronous}%
	\BibitemOpen
	\bibfield  {author} {\bibinfo {author} {\bibfnamefont {P.}~\bibnamefont
			{Xue}}, \bibinfo {author} {\bibfnamefont {L.}~\bibnamefont {Xiao}}, \bibinfo
		{author} {\bibfnamefont {G.}~\bibnamefont {Ruffolo}}, \bibinfo {author}
		{\bibfnamefont {A.}~\bibnamefont {Mazzari}}, \bibinfo {author} {\bibfnamefont
			{T.}~\bibnamefont {Temistocles}}, \bibinfo {author} {\bibfnamefont {M.~T.}\
			\bibnamefont {Cunha}},\ and\ \bibinfo {author} {\bibfnamefont
			{R.}~\bibnamefont {Rabelo}},\ }\href@noop {} {\bibfield  {journal} {\bibinfo
			{journal} {Physical Review Letters}\ }\textbf {\bibinfo {volume} {130}},\
		\bibinfo {pages} {040201} (\bibinfo {year} {2023})}\BibitemShut {NoStop}%
	\bibitem [{\citenamefont {Cabello}(2021)}]{cabello2021converting}%
	\BibitemOpen
	\bibfield  {author} {\bibinfo {author} {\bibfnamefont {A.}~\bibnamefont
			{Cabello}},\ }\href@noop {} {\bibfield  {journal} {\bibinfo  {journal}
			{Physical Review Letters}\ }\textbf {\bibinfo {volume} {127}},\ \bibinfo
		{pages} {070401} (\bibinfo {year} {2021})}\BibitemShut {NoStop}%
	\bibitem [{\citenamefont {Kirchmair}\ \emph {et~al.}(2009)\citenamefont
		{Kirchmair}, \citenamefont {Z{\"a}hringer}, \citenamefont {Gerritsma},
		\citenamefont {Kleinmann}, \citenamefont {G{\"u}hne}, \citenamefont
		{Cabello}, \citenamefont {Blatt},\ and\ \citenamefont
		{Roos}}]{kirchmair2009state}%
	\BibitemOpen
	\bibfield  {author} {\bibinfo {author} {\bibfnamefont {G.}~\bibnamefont
			{Kirchmair}}, \bibinfo {author} {\bibfnamefont {F.}~\bibnamefont
			{Z{\"a}hringer}}, \bibinfo {author} {\bibfnamefont {R.}~\bibnamefont
			{Gerritsma}}, \bibinfo {author} {\bibfnamefont {M.}~\bibnamefont
			{Kleinmann}}, \bibinfo {author} {\bibfnamefont {O.}~\bibnamefont
			{G{\"u}hne}}, \bibinfo {author} {\bibfnamefont {A.}~\bibnamefont {Cabello}},
		\bibinfo {author} {\bibfnamefont {R.}~\bibnamefont {Blatt}},\ and\ \bibinfo
		{author} {\bibfnamefont {C.~F.}\ \bibnamefont {Roos}},\ }\href@noop {}
	{\bibfield  {journal} {\bibinfo  {journal} {Nature}\ }\textbf {\bibinfo
			{volume} {460}},\ \bibinfo {pages} {494} (\bibinfo {year}
		{2009})}\BibitemShut {NoStop}%
	\bibitem [{\citenamefont {G{\"u}hne}\ \emph {et~al.}(2010)\citenamefont
		{G{\"u}hne}, \citenamefont {Kleinmann}, \citenamefont {Cabello},
		\citenamefont {Larsson}, \citenamefont {Kirchmair}, \citenamefont
		{Z{\"a}hringer}, \citenamefont {Gerritsma},\ and\ \citenamefont
		{Roos}}]{guhne2010compatibility}%
	\BibitemOpen
	\bibfield  {author} {\bibinfo {author} {\bibfnamefont {O.}~\bibnamefont
			{G{\"u}hne}}, \bibinfo {author} {\bibfnamefont {M.}~\bibnamefont
			{Kleinmann}}, \bibinfo {author} {\bibfnamefont {A.}~\bibnamefont {Cabello}},
		\bibinfo {author} {\bibfnamefont {J.-{\AA}.}\ \bibnamefont {Larsson}},
		\bibinfo {author} {\bibfnamefont {G.}~\bibnamefont {Kirchmair}}, \bibinfo
		{author} {\bibfnamefont {F.}~\bibnamefont {Z{\"a}hringer}}, \bibinfo {author}
		{\bibfnamefont {R.}~\bibnamefont {Gerritsma}},\ and\ \bibinfo {author}
		{\bibfnamefont {C.~F.}\ \bibnamefont {Roos}},\ }\href@noop {} {\bibfield
		{journal} {\bibinfo  {journal} {Physical Review A}\ }\textbf {\bibinfo
			{volume} {81}},\ \bibinfo {pages} {022121} (\bibinfo {year}
		{2010})}\BibitemShut {NoStop}%
	\bibitem [{\citenamefont {Amselem}\ \emph {et~al.}(2009)\citenamefont
		{Amselem}, \citenamefont {R{\aa}dmark}, \citenamefont {Bourennane},\ and\
		\citenamefont {Cabello}}]{amselem2009state}%
	\BibitemOpen
	\bibfield  {author} {\bibinfo {author} {\bibfnamefont {E.}~\bibnamefont
			{Amselem}}, \bibinfo {author} {\bibfnamefont {M.}~\bibnamefont
			{R{\aa}dmark}}, \bibinfo {author} {\bibfnamefont {M.}~\bibnamefont
			{Bourennane}},\ and\ \bibinfo {author} {\bibfnamefont {A.}~\bibnamefont
			{Cabello}},\ }\href@noop {} {\bibfield  {journal} {\bibinfo  {journal}
			{Physical review letters}\ }\textbf {\bibinfo {volume} {103}},\ \bibinfo
		{pages} {160405} (\bibinfo {year} {2009})}\BibitemShut {NoStop}%
	\bibitem [{\citenamefont {Lapkiewicz}\ \emph {et~al.}(2011)\citenamefont
		{Lapkiewicz}, \citenamefont {Li}, \citenamefont {Schaeff}, \citenamefont
		{Langford}, \citenamefont {Ramelow}, \citenamefont {Wie{\'s}niak},\ and\
		\citenamefont {Zeilinger}}]{lapkiewicz2011experimental}%
	\BibitemOpen
	\bibfield  {author} {\bibinfo {author} {\bibfnamefont {R.}~\bibnamefont
			{Lapkiewicz}}, \bibinfo {author} {\bibfnamefont {P.}~\bibnamefont {Li}},
		\bibinfo {author} {\bibfnamefont {C.}~\bibnamefont {Schaeff}}, \bibinfo
		{author} {\bibfnamefont {N.~K.}\ \bibnamefont {Langford}}, \bibinfo {author}
		{\bibfnamefont {S.}~\bibnamefont {Ramelow}}, \bibinfo {author} {\bibfnamefont
			{M.}~\bibnamefont {Wie{\'s}niak}},\ and\ \bibinfo {author} {\bibfnamefont
			{A.}~\bibnamefont {Zeilinger}},\ }\href@noop {} {\bibfield  {journal}
		{\bibinfo  {journal} {Nature}\ }\textbf {\bibinfo {volume} {474}},\ \bibinfo
		{pages} {490} (\bibinfo {year} {2011})}\BibitemShut {NoStop}%
	\bibitem [{\citenamefont {Szangolies}\ \emph {et~al.}(2013)\citenamefont
		{Szangolies}, \citenamefont {Kleinmann},\ and\ \citenamefont
		{G{\"u}hne}}]{szangolies2013tests}%
	\BibitemOpen
	\bibfield  {author} {\bibinfo {author} {\bibfnamefont {J.}~\bibnamefont
			{Szangolies}}, \bibinfo {author} {\bibfnamefont {M.}~\bibnamefont
			{Kleinmann}},\ and\ \bibinfo {author} {\bibfnamefont {O.}~\bibnamefont
			{G{\"u}hne}},\ }\href@noop {} {\bibfield  {journal} {\bibinfo  {journal}
			{Physical Review A-Atomic, Molecular, and Optical Physics}\ }\textbf
		{\bibinfo {volume} {87}},\ \bibinfo {pages} {050101} (\bibinfo {year}
		{2013})}\BibitemShut {NoStop}%
	\bibitem [{\citenamefont {Szangolies}(2015)}]{szangolies2015testing}%
	\BibitemOpen
	\bibfield  {author} {\bibinfo {author} {\bibfnamefont {J.}~\bibnamefont
			{Szangolies}},\ }\href@noop {} {\emph {\bibinfo {title} {Testing quantum
				contextuality: the problem of compatibility}}}\ (\bibinfo  {publisher}
	{Springer},\ \bibinfo {year} {2015})\BibitemShut {NoStop}%
	\bibitem [{\citenamefont {Vall{\'e}e}\ \emph {et~al.}(2024)\citenamefont
		{Vall{\'e}e}, \citenamefont {Emeriau}, \citenamefont {Bourdoncle},
		\citenamefont {Sohbi}, \citenamefont {Mansfield},\ and\ \citenamefont
		{Markham}}]{vallee2024corrected}%
	\BibitemOpen
	\bibfield  {author} {\bibinfo {author} {\bibfnamefont {K.}~\bibnamefont
			{Vall{\'e}e}}, \bibinfo {author} {\bibfnamefont {P.-E.}\ \bibnamefont
			{Emeriau}}, \bibinfo {author} {\bibfnamefont {B.}~\bibnamefont {Bourdoncle}},
		\bibinfo {author} {\bibfnamefont {A.}~\bibnamefont {Sohbi}}, \bibinfo
		{author} {\bibfnamefont {S.}~\bibnamefont {Mansfield}},\ and\ \bibinfo
		{author} {\bibfnamefont {D.}~\bibnamefont {Markham}},\ }\href@noop {}
	{\bibfield  {journal} {\bibinfo  {journal} {Philosophical Transactions of the
				Royal Society A}\ }\textbf {\bibinfo {volume} {382}},\ \bibinfo {pages}
		{20230011} (\bibinfo {year} {2024})}\BibitemShut {NoStop}%
	\bibitem [{\citenamefont {Spekkens}(2014)}]{spekkens2014status}%
	\BibitemOpen
	\bibfield  {author} {\bibinfo {author} {\bibfnamefont {R.~W.}\ \bibnamefont
			{Spekkens}},\ }\href@noop {} {\bibfield  {journal} {\bibinfo  {journal}
			{Foundations of Physics}\ }\textbf {\bibinfo {volume} {44}},\ \bibinfo
		{pages} {1125} (\bibinfo {year} {2014})}\BibitemShut {NoStop}%
	\bibitem [{\citenamefont {Kunjwal}\ and\ \citenamefont
		{Spekkens}(2015)}]{kunjwal2015kochen}%
	\BibitemOpen
	\bibfield  {author} {\bibinfo {author} {\bibfnamefont {R.}~\bibnamefont
			{Kunjwal}}\ and\ \bibinfo {author} {\bibfnamefont {R.~W.}\ \bibnamefont
			{Spekkens}},\ }\href@noop {} {\bibfield  {journal} {\bibinfo  {journal}
			{Physical review letters}\ }\textbf {\bibinfo {volume} {115}},\ \bibinfo
		{pages} {110403} (\bibinfo {year} {2015})}\BibitemShut {NoStop}%
	\bibitem [{\citenamefont {Krishna}\ \emph {et~al.}(2017)\citenamefont
		{Krishna}, \citenamefont {Spekkens},\ and\ \citenamefont
		{Wolfe}}]{krishna2017deriving}%
	\BibitemOpen
	\bibfield  {author} {\bibinfo {author} {\bibfnamefont {A.}~\bibnamefont
			{Krishna}}, \bibinfo {author} {\bibfnamefont {R.~W.}\ \bibnamefont
			{Spekkens}},\ and\ \bibinfo {author} {\bibfnamefont {E.}~\bibnamefont
			{Wolfe}},\ }\href@noop {} {\bibfield  {journal} {\bibinfo  {journal} {New
				Journal of Physics}\ }\textbf {\bibinfo {volume} {19}},\ \bibinfo {pages}
		{123031} (\bibinfo {year} {2017})}\BibitemShut {NoStop}%
	\bibitem [{sup()}]{supp}%
	\BibitemOpen
	\href@noop {} {}\bibinfo {howpublished}
	{\url{URL_will_be_inserted_by_publisher}}\BibitemShut {NoStop}%
	\bibitem [{\citenamefont {Kochen}\ and\ \citenamefont
		{Specker}(1990)}]{kochen1990problem}%
	\BibitemOpen
	\bibfield  {author} {\bibinfo {author} {\bibfnamefont {S.}~\bibnamefont
			{Kochen}}\ and\ \bibinfo {author} {\bibfnamefont {E.~P.}\ \bibnamefont
			{Specker}},\ }\href@noop {} {\bibfield  {journal} {\bibinfo  {journal} {Ernst
				Specker Selecta}\ ,\ \bibinfo {pages} {235}} (\bibinfo {year}
		{1990})}\BibitemShut {NoStop}%
	\bibitem [{\citenamefont {Yu}\ and\ \citenamefont {Oh}(2012)}]{Yu2011}%
	\BibitemOpen
	\bibfield  {author} {\bibinfo {author} {\bibfnamefont {S.}~\bibnamefont
			{Yu}}\ and\ \bibinfo {author} {\bibfnamefont {C.~H.}\ \bibnamefont {Oh}},\
	}\href {https://doi.org/10.1103/PhysRevLett.108.030402} {\bibfield  {journal}
		{\bibinfo  {journal} {Phys. Rev. Lett.}\ }\textbf {\bibinfo {volume} {108}},\
		\bibinfo {pages} {030402} (\bibinfo {year} {2012})}\BibitemShut {NoStop}%
	\bibitem [{\citenamefont {Cabello}\ \emph {et~al.}(1996)\citenamefont
		{Cabello}, \citenamefont {Estebaranz},\ and\ \citenamefont
		{Garc{\'\i}a-Alcaine}}]{cabello1996bell}%
	\BibitemOpen
	\bibfield  {author} {\bibinfo {author} {\bibfnamefont {A.}~\bibnamefont
			{Cabello}}, \bibinfo {author} {\bibfnamefont {J.}~\bibnamefont
			{Estebaranz}},\ and\ \bibinfo {author} {\bibfnamefont {G.}~\bibnamefont
			{Garc{\'\i}a-Alcaine}},\ }\href@noop {} {\bibfield  {journal} {\bibinfo
			{journal} {Physics Letters A}\ }\textbf {\bibinfo {volume} {212}},\ \bibinfo
		{pages} {183} (\bibinfo {year} {1996})}\BibitemShut {NoStop}%
	\bibitem [{\citenamefont {Lison{\v{e}}k}\ \emph {et~al.}(2014)\citenamefont
		{Lison{\v{e}}k}, \citenamefont {Badziag}, \citenamefont {Portillo},\ and\
		\citenamefont {Cabello}}]{lisonvek2014kochen}%
	\BibitemOpen
	\bibfield  {author} {\bibinfo {author} {\bibfnamefont {P.}~\bibnamefont
			{Lison{\v{e}}k}}, \bibinfo {author} {\bibfnamefont {P.}~\bibnamefont
			{Badziag}}, \bibinfo {author} {\bibfnamefont {J.~R.}\ \bibnamefont
			{Portillo}},\ and\ \bibinfo {author} {\bibfnamefont {A.}~\bibnamefont
			{Cabello}},\ }\href@noop {} {\bibfield  {journal} {\bibinfo  {journal}
			{Physical Review A}\ }\textbf {\bibinfo {volume} {89}},\ \bibinfo {pages}
		{042101} (\bibinfo {year} {2014})}\BibitemShut {NoStop}%
	\bibitem [{\citenamefont {Mair}\ \emph {et~al.}(2001)\citenamefont {Mair},
		\citenamefont {Vaziri}, \citenamefont {Weihs},\ and\ \citenamefont
		{Zeilinger}}]{mair2001entanglement}%
	\BibitemOpen
	\bibfield  {author} {\bibinfo {author} {\bibfnamefont {A.}~\bibnamefont
			{Mair}}, \bibinfo {author} {\bibfnamefont {A.}~\bibnamefont {Vaziri}},
		\bibinfo {author} {\bibfnamefont {G.}~\bibnamefont {Weihs}},\ and\ \bibinfo
		{author} {\bibfnamefont {A.}~\bibnamefont {Zeilinger}},\ }\href@noop {}
	{\bibfield  {journal} {\bibinfo  {journal} {Nature}\ }\textbf {\bibinfo
			{volume} {412}},\ \bibinfo {pages} {313} (\bibinfo {year}
		{2001})}\BibitemShut {NoStop}%
	\bibitem [{\citenamefont {Torres}\ \emph {et~al.}(2003)\citenamefont {Torres},
		\citenamefont {Alexandrescu},\ and\ \citenamefont
		{Torner}}]{torres2003quantum}%
	\BibitemOpen
	\bibfield  {author} {\bibinfo {author} {\bibfnamefont {J.~P.}\ \bibnamefont
			{Torres}}, \bibinfo {author} {\bibfnamefont {A.}~\bibnamefont
			{Alexandrescu}},\ and\ \bibinfo {author} {\bibfnamefont {L.}~\bibnamefont
			{Torner}},\ }\href {https://doi.org/10.1103/PhysRevA.68.050301} {\bibfield
		{journal} {\bibinfo  {journal} {Physical Review A}\ }\textbf {\bibinfo
			{volume} {68}},\ \bibinfo {pages} {050301} (\bibinfo {year}
		{2003})}\BibitemShut {NoStop}%
	\bibitem [{\citenamefont {Leach}\ \emph {et~al.}(2010)\citenamefont {Leach},
		\citenamefont {Jack}, \citenamefont {Romero}, \citenamefont {Jha},
		\citenamefont {Yao}, \citenamefont {Franke-Arnold}, \citenamefont {Ireland},
		\citenamefont {Boyd}, \citenamefont {Barnett},\ and\ \citenamefont
		{Padgett}}]{leach2010quantum}%
	\BibitemOpen
	\bibfield  {author} {\bibinfo {author} {\bibfnamefont {J.}~\bibnamefont
			{Leach}}, \bibinfo {author} {\bibfnamefont {B.}~\bibnamefont {Jack}},
		\bibinfo {author} {\bibfnamefont {J.}~\bibnamefont {Romero}}, \bibinfo
		{author} {\bibfnamefont {A.~K.}\ \bibnamefont {Jha}}, \bibinfo {author}
		{\bibfnamefont {A.~M.}\ \bibnamefont {Yao}}, \bibinfo {author} {\bibfnamefont
			{S.}~\bibnamefont {Franke-Arnold}}, \bibinfo {author} {\bibfnamefont {D.~G.}\
			\bibnamefont {Ireland}}, \bibinfo {author} {\bibfnamefont {R.~W.}\
			\bibnamefont {Boyd}}, \bibinfo {author} {\bibfnamefont {S.~M.}\ \bibnamefont
			{Barnett}},\ and\ \bibinfo {author} {\bibfnamefont {M.~J.}\ \bibnamefont
			{Padgett}},\ }\href@noop {} {\bibfield  {journal} {\bibinfo  {journal}
			{Science}\ }\textbf {\bibinfo {volume} {329}},\ \bibinfo {pages} {662}
		(\bibinfo {year} {2010})}\BibitemShut {NoStop}%
	\bibitem [{\citenamefont {Erhard}\ \emph {et~al.}(2018)\citenamefont {Erhard},
		\citenamefont {Fickler}, \citenamefont {Krenn},\ and\ \citenamefont
		{Zeilinger}}]{erhard2018twisted}%
	\BibitemOpen
	\bibfield  {author} {\bibinfo {author} {\bibfnamefont {M.}~\bibnamefont
			{Erhard}}, \bibinfo {author} {\bibfnamefont {R.}~\bibnamefont {Fickler}},
		\bibinfo {author} {\bibfnamefont {M.}~\bibnamefont {Krenn}},\ and\ \bibinfo
		{author} {\bibfnamefont {A.}~\bibnamefont {Zeilinger}},\ }\href@noop {}
	{\bibfield  {journal} {\bibinfo  {journal} {Light: Science \& Applications}\
		}\textbf {\bibinfo {volume} {7}},\ \bibinfo {pages} {17146} (\bibinfo {year}
		{2018})}\BibitemShut {NoStop}%
	\bibitem [{\citenamefont {Padgett}\ \emph {et~al.}(2004)\citenamefont
		{Padgett}, \citenamefont {Courtial},\ and\ \citenamefont
		{Allen}}]{padgett2004light}%
	\BibitemOpen
	\bibfield  {author} {\bibinfo {author} {\bibfnamefont {M.}~\bibnamefont
			{Padgett}}, \bibinfo {author} {\bibfnamefont {J.}~\bibnamefont {Courtial}},\
		and\ \bibinfo {author} {\bibfnamefont {L.}~\bibnamefont {Allen}},\
	}\href@noop {} {\bibfield  {journal} {\bibinfo  {journal} {Physics today}\
		}\textbf {\bibinfo {volume} {57}},\ \bibinfo {pages} {35} (\bibinfo {year}
		{2004})}\BibitemShut {NoStop}%
	\bibitem [{\citenamefont {Molina-Terriza}\ \emph {et~al.}(2001)\citenamefont
		{Molina-Terriza}, \citenamefont {Torres},\ and\ \citenamefont
		{Torner}}]{molina2001management}%
	\BibitemOpen
	\bibfield  {author} {\bibinfo {author} {\bibfnamefont {G.}~\bibnamefont
			{Molina-Terriza}}, \bibinfo {author} {\bibfnamefont {J.~P.}\ \bibnamefont
			{Torres}},\ and\ \bibinfo {author} {\bibfnamefont {L.}~\bibnamefont
			{Torner}},\ }\href@noop {} {\bibfield  {journal} {\bibinfo  {journal}
			{Physical review letters}\ }\textbf {\bibinfo {volume} {88}},\ \bibinfo
		{pages} {013601} (\bibinfo {year} {2001})}\BibitemShut {NoStop}%
	\bibitem [{\citenamefont {Walborn}\ \emph {et~al.}(2004)\citenamefont
		{Walborn}, \citenamefont {De~Oliveira}, \citenamefont {Thebaldi},\ and\
		\citenamefont {Monken}}]{walborn2004entanglement}%
	\BibitemOpen
	\bibfield  {author} {\bibinfo {author} {\bibfnamefont {S.}~\bibnamefont
			{Walborn}}, \bibinfo {author} {\bibfnamefont {A.}~\bibnamefont
			{De~Oliveira}}, \bibinfo {author} {\bibfnamefont {R.}~\bibnamefont
			{Thebaldi}},\ and\ \bibinfo {author} {\bibfnamefont {C.}~\bibnamefont
			{Monken}},\ }\href@noop {} {\bibfield  {journal} {\bibinfo  {journal}
			{Physical Review A}\ }\textbf {\bibinfo {volume} {69}},\ \bibinfo {pages}
		{023811} (\bibinfo {year} {2004})}\BibitemShut {NoStop}%
	\bibitem [{\citenamefont {Leach}\ \emph {et~al.}(2005)\citenamefont {Leach},
		\citenamefont {Dennis}, \citenamefont {Courtial},\ and\ \citenamefont
		{Padgett}}]{leach2005vortex}%
	\BibitemOpen
	\bibfield  {author} {\bibinfo {author} {\bibfnamefont {J.}~\bibnamefont
			{Leach}}, \bibinfo {author} {\bibfnamefont {M.~R.}\ \bibnamefont {Dennis}},
		\bibinfo {author} {\bibfnamefont {J.}~\bibnamefont {Courtial}},\ and\
		\bibinfo {author} {\bibfnamefont {M.~J.}\ \bibnamefont {Padgett}},\
	}\href@noop {} {\bibfield  {journal} {\bibinfo  {journal} {New Journal of
				Physics}\ }\textbf {\bibinfo {volume} {7}},\ \bibinfo {pages} {55} (\bibinfo
		{year} {2005})}\BibitemShut {NoStop}%
	\bibitem [{\citenamefont {Bennett}\ \emph {et~al.}(1996)\citenamefont
		{Bennett}, \citenamefont {Bernstein}, \citenamefont {Popescu},\ and\
		\citenamefont {Schumacher}}]{bennett1996concentrating}%
	\BibitemOpen
	\bibfield  {author} {\bibinfo {author} {\bibfnamefont {C.~H.}\ \bibnamefont
			{Bennett}}, \bibinfo {author} {\bibfnamefont {H.~J.}\ \bibnamefont
			{Bernstein}}, \bibinfo {author} {\bibfnamefont {S.}~\bibnamefont {Popescu}},\
		and\ \bibinfo {author} {\bibfnamefont {B.}~\bibnamefont {Schumacher}},\
	}\href@noop {} {\bibfield  {journal} {\bibinfo  {journal} {Physical Review
				A}\ }\textbf {\bibinfo {volume} {53}},\ \bibinfo {pages} {2046} (\bibinfo
		{year} {1996})}\BibitemShut {NoStop}%
	\bibitem [{\citenamefont {Law}\ and\ \citenamefont
		{Eberly}(2004)}]{law2004analysis}%
	\BibitemOpen
	\bibfield  {author} {\bibinfo {author} {\bibfnamefont {C.}~\bibnamefont
			{Law}}\ and\ \bibinfo {author} {\bibfnamefont {J.}~\bibnamefont {Eberly}},\
	}\href@noop {} {\bibfield  {journal} {\bibinfo  {journal} {Physical review
				letters}\ }\textbf {\bibinfo {volume} {92}},\ \bibinfo {pages} {127903}
		(\bibinfo {year} {2004})}\BibitemShut {NoStop}%
	\bibitem [{\citenamefont {Dada}\ \emph {et~al.}(2011)\citenamefont {Dada},
		\citenamefont {Leach}, \citenamefont {Buller}, \citenamefont {Padgett},\ and\
		\citenamefont {Andersson}}]{dada2011experimental}%
	\BibitemOpen
	\bibfield  {author} {\bibinfo {author} {\bibfnamefont {A.~C.}\ \bibnamefont
			{Dada}}, \bibinfo {author} {\bibfnamefont {J.}~\bibnamefont {Leach}},
		\bibinfo {author} {\bibfnamefont {G.~S.}\ \bibnamefont {Buller}}, \bibinfo
		{author} {\bibfnamefont {M.~J.}\ \bibnamefont {Padgett}},\ and\ \bibinfo
		{author} {\bibfnamefont {E.}~\bibnamefont {Andersson}},\ }\href@noop {}
	{\bibfield  {journal} {\bibinfo  {journal} {Nature Physics}\ }\textbf
		{\bibinfo {volume} {7}},\ \bibinfo {pages} {677} (\bibinfo {year}
		{2011})}\BibitemShut {NoStop}%
	\bibitem [{\citenamefont {Cabello}\ \emph {et~al.}(2014)\citenamefont
		{Cabello}, \citenamefont {Severini},\ and\ \citenamefont
		{Winter}}]{cabello2014graph}%
	\BibitemOpen
	\bibfield  {author} {\bibinfo {author} {\bibfnamefont {A.}~\bibnamefont
			{Cabello}}, \bibinfo {author} {\bibfnamefont {S.}~\bibnamefont {Severini}},\
		and\ \bibinfo {author} {\bibfnamefont {A.}~\bibnamefont {Winter}},\
	}\href@noop {} {\bibfield  {journal} {\bibinfo  {journal} {Physical review
				letters}\ }\textbf {\bibinfo {volume} {112}},\ \bibinfo {pages} {040401}
		(\bibinfo {year} {2014})}\BibitemShut {NoStop}%
	\bibitem [{\citenamefont {Lov{\'a}sz}(1979)}]{lovasz1979shannon}%
	\BibitemOpen
	\bibfield  {author} {\bibinfo {author} {\bibfnamefont {L.}~\bibnamefont
			{Lov{\'a}sz}},\ }\href@noop {} {\bibfield  {journal} {\bibinfo  {journal}
			{IEEE Transactions on Information theory}\ }\textbf {\bibinfo {volume}
			{25}},\ \bibinfo {pages} {1} (\bibinfo {year} {1979})}\BibitemShut {NoStop}%
	\bibitem [{\citenamefont {Wang}\ \emph {et~al.}(2022)\citenamefont {Wang},
		\citenamefont {Zhang}, \citenamefont {Luan}, \citenamefont {Um},
		\citenamefont {Wang}, \citenamefont {Qiao}, \citenamefont {Xie},
		\citenamefont {Zhang}, \citenamefont {Cabello},\ and\ \citenamefont
		{Kim}}]{wang2022significant}%
	\BibitemOpen
	\bibfield  {author} {\bibinfo {author} {\bibfnamefont {P.}~\bibnamefont
			{Wang}}, \bibinfo {author} {\bibfnamefont {J.}~\bibnamefont {Zhang}},
		\bibinfo {author} {\bibfnamefont {C.-Y.}\ \bibnamefont {Luan}}, \bibinfo
		{author} {\bibfnamefont {M.}~\bibnamefont {Um}}, \bibinfo {author}
		{\bibfnamefont {Y.}~\bibnamefont {Wang}}, \bibinfo {author} {\bibfnamefont
			{M.}~\bibnamefont {Qiao}}, \bibinfo {author} {\bibfnamefont {T.}~\bibnamefont
			{Xie}}, \bibinfo {author} {\bibfnamefont {J.-N.}\ \bibnamefont {Zhang}},
		\bibinfo {author} {\bibfnamefont {A.}~\bibnamefont {Cabello}},\ and\ \bibinfo
		{author} {\bibfnamefont {K.}~\bibnamefont {Kim}},\ }\href@noop {} {\bibfield
		{journal} {\bibinfo  {journal} {Science Advances}\ }\textbf {\bibinfo
			{volume} {8}},\ \bibinfo {pages} {eabk1660} (\bibinfo {year}
		{2022})}\BibitemShut {NoStop}%
	\bibitem [{\citenamefont {Hu}\ \emph {et~al.}(2023)\citenamefont {Hu},
		\citenamefont {Xie}, \citenamefont {Arora}, \citenamefont {Ai}, \citenamefont
		{Bharti}, \citenamefont {Zhang}, \citenamefont {Wu}, \citenamefont {Chen},
		\citenamefont {Cui}, \citenamefont {Liu} \emph {et~al.}}]{hu2023self}%
	\BibitemOpen
	\bibfield  {author} {\bibinfo {author} {\bibfnamefont {X.-M.}\ \bibnamefont
			{Hu}}, \bibinfo {author} {\bibfnamefont {Y.}~\bibnamefont {Xie}}, \bibinfo
		{author} {\bibfnamefont {A.~S.}\ \bibnamefont {Arora}}, \bibinfo {author}
		{\bibfnamefont {M.-Z.}\ \bibnamefont {Ai}}, \bibinfo {author} {\bibfnamefont
			{K.}~\bibnamefont {Bharti}}, \bibinfo {author} {\bibfnamefont
			{J.}~\bibnamefont {Zhang}}, \bibinfo {author} {\bibfnamefont
			{W.}~\bibnamefont {Wu}}, \bibinfo {author} {\bibfnamefont {P.-X.}\
			\bibnamefont {Chen}}, \bibinfo {author} {\bibfnamefont {J.-M.}\ \bibnamefont
			{Cui}}, \bibinfo {author} {\bibfnamefont {B.-H.}\ \bibnamefont {Liu}}, \emph
		{et~al.},\ }\href@noop {} {\bibfield  {journal} {\bibinfo  {journal} {npj
				Quantum Information}\ }\textbf {\bibinfo {volume} {9}},\ \bibinfo {pages}
		{103} (\bibinfo {year} {2023})}\BibitemShut {NoStop}%
\end{thebibliography}

\end{document}